\documentstyle[aps,floats,prl,psfig]{revtex}

\newcommand{\ApJ}{Astrophys. J.}
\newcommand{\PRL}{Phys. Rev. Lett.}
\newcommand{\PRD}{Phys. Rev. D}
\newcommand{\MNRAS}{MNRAS}

\renewcommand{\l}{{\bf l}}
\newcommand{\la}{{{\bf l}_1}}

\newcommand{\da}{d_A}

\newcommand{\tot}{{\rm t}}

\newcommand{\vecl}{{\bf l}}
\newcommand{\vecla}{{{\bf l}_1}}

\newcommand{\intl}[1]{\int {d^2 {\bf l}_#1 \over (2\pi)^2}}

\newlength{\tskip}
\newlength{\colwidth}

\newcommand{\beq}{\begin{equation}}
\newcommand{\eeq}{\end{equation}}
\newcommand{\beqa}{\begin{eqnarray}}
\newcommand{\eeqa}{\end{eqnarray}}
\newcommand{\bn}{\hat{\bf n}}
\newcommand{\bm}{\hat{\bf m}}

\newcommand{\rad}{r}    % comoving radial distance

\newcommand{\len}{\phi}

\begin{document}
%@twocolumnfalse\endcsname

\title{A Lensing Reconstruction of Primordial Cosmic Microwave Background Polarization}
\author{Asantha Cooray}
\address{
Theoretical Astrophysics, California Institute of Technology,
Pasadena, California 91125\\}

%\date{To be submitted to Phys. Rev. D. --- October 2001}

\maketitle

%------------------------------------------------------------------------------

\begin{abstract}
We discuss a possibility to directly reconstruct the CMB  polarization field at the
last scattering surface by accounting for modifications imposed by the gravitational lensing
effect. The suggested method requires a tracer field of the large scale structure lensing potentials that
deflected propagating CMB photons from the last scattering surface. This required information can come 
from a variety of observations on the large scale structure matter distribution, including convergence
reconstructed from lensing shear studies involving galaxy shapes.  In the case of so-called curl, or B,-modes of
CMB polarization, the reconstruction allows one to identify the distinct signature of inflationary gravitational waves.
\end{abstract}

%]

%------------------------------------------------------------------------------
% User-supplied List of keywords.

%\pacs{PACS numbers: 98.80.Es,95.85.Nv,98.35.Ce,98.70.Vc
%\hfill}
%]

%------------------------------------------------------------------------------

\section{Introduction}

The advent of high sensitivity cosmic microwave background (CMB) experiments will soon allow detailed
observational studies on the CMB polarization field. In addition to a confirmation of
the basic picture on how anisotropies were generated during the recombination era \cite{PeeYu70}, 
CMB polarization observations, at large angular scales, may allow a measurement of the reionization redshift based on
excess power created during the rescattering of CMB quadrupole 
at the reionization surface \cite{Zal97}. An additional use of polarization observations is the 
potential detection of inflationary gravitational waves 
based on its contribution to the so-called curl or B-modes of polarization \cite{KamKosSte97}. 

In addition to primordial contributions involving density perturbations, which contribute solely to E-modes, and
tensor modes or gravitational waves, the polarization field one observes today also contains secondary contributions 
associated with the large scale structure. This is similar to well known secondary contributions that
dominate the small scale temperature anisotropy signal \cite{Coo02}. 
In the case of B-modes, the arcminute scale polarization is dominated by
a fractional conversion of the dominant E-mode contribution via the gravitational lensing angular deflection of CMB photons
\cite{ZalSel98}. In Ref. \cite{Kesetal02}, we discussed an approach to separate the gravitational-wave signature in 
B-modes from the dominant, and contaminant, weak lensing contribution at arcminute scales. We also investigated 
limits on gravitational wave B-mode detection after a model-independent lensing substraction. 
This calculation followed previous discussions in the literature on how  CMB data can be used for an extraction of 
statistics involving the gravitational lensing effect, such as shear or deflection angles \cite{SelZal99,Hu01,CooKes02}.
 
In general, the reconstruction of lensing deflections from CMB data require a priori information on the
primordial CMB contribution at the last scattering surface. This knowledge cannot easily be obtained from data alone
since the CMB contribution measured today involve contributions resulting from secondary effects and any modifications
during the transit to us from the last scattering surface.
On the other hand, we can consider an alternative approach to an analogous problem involving
a reconstruction of the primordial CMB contribution, at the last scattering surface, by accounting for
modifications due to large scale structure. This is, effectively, an
inverse problem from what was discussed in \cite{Hu01,CooKes02} and require prior information
related to the lensing effect on CMB, such as the mass distribution of large scale structure in which photons propagated. 
In the present study, we briefly discuss this possibility using a tracer field of the
large scale structure potentials that deflected CMB photons via the gravitational lensing effect.

The discussion presented here follows the recent work of Ref. \cite{CooKes02} where we discussed the lensing
reconstruction from CMB data given information related to primordial CMB anisotropy and polarization fields. 
We refer the reader to the above paper for basic details of lensing and other important
ingredients related to the calculation. To illustrate our results, we assume the currently favored $\Lambda$CDM cosmology.

\section{Calculation}

The lensing effect on CMB data can be described simply as a transfer of power via changes to 
photon propagation directions on the sky.   
In Fourier space, we can consider the E- and B-mode decomposition introduced in Ref. \cite{KamKosSte97}
such that ${}_\pm Y(\l) = E(\l)\pm i B(\l)$. We denote the primordial contribution at the
last scattering surface with a $\tilde Y$, while the polarization field observed today is denoted with $Y$. 
We can consider the lensing modification following Ref. \cite{Hu00} and write
\begin{eqnarray}
{}_\pm Y(\l) &=& {}_\pm \tilde Y(\l)
-
\intl{1}
{}_\pm \tilde Y(\la)
 e^{\pm 2i (\varphi_{\vecl_1}- \varphi_{\vecl})} L(\l,\l_1) \,,
\end{eqnarray}
where
\begin{eqnarray}
\label{eqn:lfactor}
L(\vecla,\vecla') &=& \len(\vecla-\vecla') \, (\vecla - \vecla') \cdot \vecla'
+{1 \over 2} \intl{1''} \len(\vecla'') \len^*(\vecla'' + \vecla' - \vecla) \, (\vecla'' \cdot \vecla')
                (\vecla'' + \vecla' - \vecla)\cdot
                             \vecla' \,,  \nonumber
\end{eqnarray}
to the second order in the projected lensing potential $\phi$ given by \cite{CooKes02,Hu00}
\begin{eqnarray}
\phi(\bm)&=&- 2 \int_0^{\rad_0} d\rad
\frac{\da(\rad_0-\rad)}{\da(\rad)\da(\rad_0)}
\Phi (\rad,\hat{{\bf m}}\rad ) \,.
\label{eqn:lenspotential}
\end{eqnarray}
Here, $\rad$ is the conformal time or comoving radial distance, $\da$ is the analogous comoving angular diameter
distance, $\Phi$ is the gravitational potential. The latter is related to  fluctuations in the density field
via the Poisson equation. The deflection angle associated with lensing is the gradient of the projected potential,
$\alpha(\bn) = \nabla \phi(\bn)$.
Using deflections to the first order, we write
\begin{eqnarray}
E(\vecl) &=& \tilde E(\vecl) - \intl{1} \Big[\tilde E(\vecl_1) \cos 2 (\varphi_{\vecl_1} -\varphi_\vecl)  - \tilde B(\vecl_1) \sin 2 (\varphi_{\vecl_1} -\varphi_\vecl)\Big] \phi(\vecl-\vecl_1) (\vecl-\vecl_1) \cdot \vecl_1 \nonumber \\
B(\vecl) &=& \tilde B(\vecl) - \intl{1} \Big[\tilde E(\vecl_1) \sin 2 (\varphi_{\vecl_1} -\varphi_\vecl) + \tilde B(\vecl_1) \cos 2 (\varphi_{\vecl_1} -\varphi_\vecl)\Big] \phi(\vecl-\vecl_1) (\vecl-\vecl_1) \cdot \vecl_1 \, .
\label{eqn:pollens}
\end{eqnarray}

In order to extract the primordial contribution, unaffected by lensing, consider the combination of $E$ and $B$ fields
with a tracer field of the deflecting potential, which we label here as $X$. This tracer map has the
property that it does not correlate strongly with either $\tilde E$ or $\tilde B$, the quantities at the
last scattering. Thus, only correlation expected is the one resulting due to lensing, and to a lesser extent
from other secondary contributions to polarization. Since latter contributions involving scattering
effects associated with large scale structure is significantly smaller \cite{Hu99}, we will ignore them in this
discussion. Additionally, secondary scattering effects associated with polarization can be removed from data based on their
spectral dependences. This is similar to the approach suggested in
Ref. \cite{Cooetal00} to remove the Sunyaev-Zel'dovich effect (SZ; \cite{SunZel80}) from dominant temperature anisotropies. 
We consider a map created by taking the product of the polarization field and the
tracer field, e.g., $EX$. In Fourier space, we  can write this product  as a convolution of the Fourier moments of the
two fields:
\begin{eqnarray}
&&(EX)(\vecl) = \frac{1}{2}\intl{1} \left[E(\vecl_1)X(\vecl-\vecl_1) + X(\vecl_1)E(\vecl-\vecl_1)\right] W(\vecl,\vecl_1)\, 
\label{eqn:expand}
\end{eqnarray}
where $W$ is a Fourier-space based filter that we will design  to maximize the reconstruction of the
primordial polarization information. Similarly, we can also construct another quadratic statistic involving  B and X-fields.
Using equation~(\ref{eqn:pollens}) in (\ref{eqn:expand}), we can simplify to obtain
\begin{eqnarray}
(EX)(\vecl) &=& \frac{1}{2}\tilde E(\vecl) \intl{1} W(\vecl,\vecl_1)
\Big[C_{|\vecl-\vecl_1|}^{X\phi}  \; (\vecl-\vecl_1) \cdot \vecl \;
\cos 2 (\varphi_{\vecl} -\varphi_{\vecl_1}) + C_{l_1}^{X\phi} \; \vecl_1 \cdot \vecl \;
\cos 2 (\varphi_{\vecl} -\varphi_{|\vecl-\vecl_1|}) \Big]\nonumber \\
&-&\frac{1}{2}\tilde B(\vecl) \intl{1} W(\vecl,\vecl_1)\Big[C_{|\vecl-\vecl_1|}^{X\phi} \; (\vecl-\vecl_1) \cdot \vecl \; \sin 2 (\varphi_{\vecl} -\varphi_{\vecl_1}) + C_{l_1}^{X\phi} \;  \vecl_1 \cdot \vecl  \;
\sin 2 (\varphi_{\vecl} -\varphi_{|\vecl-\vecl_1|}) \Big] \, ,
\label{eqn:EX}
\end{eqnarray}
while, analogous contribution to the quadratic statistic involving $B$ and $X$ fields, $(BX)(\vecl)$, is
\begin{eqnarray}
(BX)(\vecl) &=& \frac{1}{2}\tilde E(\vecl) \intl{1} W(\vecl,\vecl_1)
\Big[C_{|\vecl-\vecl_1|}^{X\phi} \; (\vecl-\vecl_1) \cdot \vecl \;
\sin 2 (\varphi_{\vecl} -\varphi_{\vecl_1}) + C_{l_1}^{X\phi}  \; \vecl_1 \cdot \vecl \;
\sin 2 (\varphi_{\vecl} -\varphi_{|\vecl-\vecl_1|}) \Big]\nonumber \\
&+&\frac{1}{2}\tilde B(\vecl) \intl{1} W(\vecl,\vecl_1)\Big[C_{|\vecl-\vecl_1|}^{X\phi} \;  (\vecl-\vecl_1) \cdot \vecl 
\; \cos 2 (\varphi_{\vecl} -\varphi_{\vecl_1}) + C_{l_1}^{X\phi} \; \vecl_1 \cdot \vecl  \;
\cos 2 (\varphi_{\vecl} -\varphi_{|\vecl-\vecl_1|}) \Big] \, .
\label{eqn:BX}
\end{eqnarray}
As written, these quadratic combinations are proportional to the primordial polarization contributions at the
last scattering surface, $\tilde E(\vecl)$ and $\tilde B(\vecl)$. This is essentially the basis for the suggested reconstruction.
In \cite{Hu01,CooKes02}, suggested quadratic statistics involve combinations of $EE$ and $EB$ such that these lead
to terms which are proportional to $\phi$ times integrals involving $C_l^{\tilde E \tilde E}$ and $C_l^{\tilde B \tilde B}$;
Thus, with some knowledge on these contributions, it was possible to reconstruct lensing deflection potentials.
Our approach, though similar, involves the inverse problem of constructing $\tilde E$ and $\tilde B$ given information related to
how $\phi$ behaves. In above, $C_l^{X\phi}$ is the
cross-correlation between tracer $X$ field and the lensing potentials $\phi$ defined such that
\begin{equation}
\langle X(\vecl) \phi(\vecl') \rangle = (2\pi)^2 \delta_D(\vecl+\vecl') C_l^{X\phi} \, .
\end{equation}
We can write this cross angular power spectrum as
\begin{equation}
C_l^{X\phi} = {2 \over \pi} \int k^2 dk P_{\delta\delta}(k)
                I_l^X(k) I_l^\phi(k) \,,
\end{equation}
where
\begin{equation}
I_l^X(k) = \int_0^{\rad_0} d\rad W^X(k,\rad) j_l(k\rad) \, ,
\end{equation}
and similarly for $I_l^\phi(k)$. Here, $W^i(k,\rad)$ is the window function of the X or $\phi$ 
in radial coordinates, $j_l$'s are the spherical Bessel functions and $P_{\delta \delta}(k)$ is the power 
spectrum of density fluctuations (see, Ref. \cite{CooKes02} for details).

To extract power spectra associated with the primordial polarization, $C_l^{\tilde E \tilde E}$ and
$C_l^{\tilde B \tilde B}$, we can either use power spectra constructed from correlating quadratic statistics themselves,
e.g., $\langle (EX)(\vecl) (EX)(\vecl') \rangle$, or the quadratic statistic correlated with the polarization field,
e.g., $\langle (EX)(\vecl) E(\vecl') \rangle$. The latter approach involves an extraction of the power spectrum from the
bispectrum formed by the fields traced by the quadratic statistic and either $E$ or $B$; this is similar
to what was suggested in Ref. \cite{Coo01} to extract the lensing-SZ correlation from squared temperature-temperature
power spectrum. The method based on $\langle (EX)(\vecl) (EX)(\vecl') \rangle$
extracts the power spectrum information via the trispectrum formed by lensing and tracer field correlation associated with 
quadratic statistics. This is analogous to techniques in Refs. \cite{Hu01,CooKes02} to extract the lensing
information using CMB data. Here, we will use the approach based on the bispectrum since it has less
noise contributions; the approach based on the trispectrum has additional non-Gaussian noise contributions which
can complicate the extraction while lowering the  associated signal-to-noise ratio  \cite{CooKes02}.

We first consider the power spectrum formed by $\langle (EX)(\vecl) E(\vecl') \rangle$. 
Following our standard definition, we define the associated power spectrum in this case as
\begin{equation}
\langle (EX)(\vecl) E(\vecl') \rangle = (2\pi)^2 \delta_D(\vecl+\vecl') C_l^{EX-E} \, .
\end{equation}
Similarly, we define the power spectra involved with other combinations of $E$ and $B$ with $(EX)$ and $(BX)$.

We can write the angular power spectrum associated with the $\langle (EX)(\vecl) E(\vecl') \rangle$ correlation as
\begin{eqnarray}
C_l^{EX-E} &=& \frac{1}{2} C_l^{\tilde E \tilde E} 
\intl{1} W(\vecl,\vecl_1) \Big[C_{|\vecl-\vecl_1|}^{X\phi}  \; (\vecl-\vecl_1) \cdot \vecl \;
\cos 2 (\varphi_{\vecl} -\varphi_{\vecl_1}) + C_{l_1}^{X\phi} \; \vecl_1 \cdot \vecl \;
\cos 2 (\varphi_{\vecl} -\varphi_{|\vecl-\vecl_1|}) \Big] \, .
\label{eq:exe}
\end{eqnarray}
Note that there is no contribution coming from the $\tilde B$ term associated with $(EX)(\vecl)$ since there is no
correlation between $\tilde E(\vecl')$ and $\tilde B(\vecl)$ due to parity considerations. 
Also, there is no resulting contribution associated with the first order lensing term of $E$ as lensing
deflections, $\phi$, should not correlate with the primordial polarization field.
Similarly, the contribution to $\langle (EX)(\vecl) B(\vecl') \rangle$ results in a power spectrum given by
\begin{eqnarray}
C_l^{EX-B} &=& -\frac{1}{2} C_l^{\tilde B \tilde B} 
\intl{1} W(\vecl,\vecl_1) \Big[C_{|\vecl-\vecl_1|}^{X\phi}  \; (\vecl-\vecl_1) \cdot \vecl \;
\sin 2 (\varphi_{\vecl} -\varphi_{\vecl_1}) + C_{l_1}^{X\phi} \; \vecl_1 \cdot \vecl \;
\sin 2 (\varphi_{\vecl} -\varphi_{|\vecl-\vecl_1|}) \Big] \, .
\label{eq:exb}
\end{eqnarray}

Since $(BX)$ provides an addition statistic to correlated with polarization maps, we can also consider the 
cross-correlation of $(BX)(\vecl)$ with $E(\vecl)$ and $B(\vecl)$. We can write the two contributions as
\begin{eqnarray}
C_l^{BX-E} &=& \frac{1}{2} C_l^{\tilde E \tilde E} 
\intl{1} W(\vecl,\vecl_1) \Big[C_{|\vecl-\vecl_1|}^{X\phi}  \; (\vecl-\vecl_1) \cdot \vecl \;
\sin 2 (\varphi_{\vecl} -\varphi_{\vecl_1}) + C_{l_1}^{X\phi} \; \vecl_1 \cdot \vecl \;
\sin 2 (\varphi_{\vecl} -\varphi_{|\vecl-\vecl_1|}) \Big] \, ,
\label{eq:bxe}
\end{eqnarray}
and
\begin{eqnarray}
C_l^{BX-B} &=& \frac{1}{2} C_l^{\tilde B \tilde B} 
\intl{1} W(\vecl,\vecl_1) \Big[C_{|\vecl-\vecl_1|}^{X\phi}  \; (\vecl-\vecl_1) \cdot \vecl \;
\cos 2 (\varphi_{\vecl} -\varphi_{\vecl_1}) + C_{l_1}^{X\phi} \; \vecl_1 \cdot \vecl \;
\cos 2 (\varphi_{\vecl} -\varphi_{|\vecl-\vecl_1|}) \Big] \, ,
\label{eq:bxb}
\end{eqnarray}
respectively.
As written, $C_l^{EX-E}$ and $C_l^{BX-E}$ are proportional to $C_l^{\tilde E \tilde E}$ while $C_l^{EX-B}$ and
$C_l^{BX-B}$ are proportional to
$C_l^{\tilde B \tilde B}$. This is the basis of the present approach which is leading to an extraction of these
primordial polarization power spectra uncontaminated by the lensing contribution.
With an appropriate normalization and a description for the filter, $W(\vecl,\vecl_1)$  that optimizes the
polarization extraction, one can effectively extract $C_l^{\tilde B \tilde B}$ and
$C_l^{\tilde E \tilde E}$ from the higher order statistics involving polarization maps today and a map of the
large scale structure.

The signal-to-noise ratio for the extraction of the primordial power spectra of polarization, for example,
in the case of $C_l^{\tilde B \tilde B}$ extracted using $\langle (BX)(\vecl) B(\vecl') \rangle$, is
\begin{eqnarray}
&&\left(\frac{\rm S}{\rm N}\right)^2	 = \nonumber \\
&&\sum_l \frac{ f_{\rm sky} (2l+1) \left[ C_l^{\tilde B \tilde B} \right]^2} {\left[C_l^{\tilde B \tilde B} \right]^2 + 2 \left[
\intl{1} W(\vecl,\vecl_1) \left\{ C_{|\vecl-\vecl_1|}^{X\phi}  \; (\vecl-\vecl_1) \cdot \vecl \;
\cos 2 (\varphi_{\vecl} -\varphi_{\vecl_1}) + C_{l_1}^{X\phi} \; \vecl_1 \cdot \vecl \;
\cos 2 (\varphi_{\vecl} -\varphi_{|\vecl-\vecl_1|}) \right\} \right]^{-2} N_l^{BX} C_l^{B,\tot} } \, ,
\end{eqnarray}
where $f_{\rm sky}$ is the fraction of sky covered and noise contributions given by $N_l^{BX}$ where
\begin{equation}
N_l^{BX} = 
\frac{1}{2}\intl{1} W^2(\vecl,\vecl_1) \left[C_{l_1}^{B,\tot} C_{|\vecl-\vecl_1|}^{X,\tot} +C_{|\vecl-\vecl_1|}^{B,\tot} C_{l_1}^{X,\tot} \right] \, ,
\end{equation}
is the noise associated with $(BX)$ statistic. In above,
$C_l^{B,\tot}$, and $C_l^{X,\tot}$, are the total contributions to  B-mode and X-fields
respectively. We can write these contributions as $C_l^{i,\tot} = C_l^i +C_l^n+C_l^s$ where $C_l^n$ is any noise
contribution, for example, instrumental noise in the B-map, and $C_l^s$ any secondary contribution, such as the
dominant lensing contribution in the B-map.

In order to maximize the signal-to-noise ratio for the extraction of primordial polarization power spectra, we
follow the approach in Ref. \cite{CooKes02} and consider a description for the filter.
The maximum signal-to-noise ratio is obtained when
\begin{eqnarray}
W(\vecl,\vecl_1) = \frac{\Big[C_{|\vecl-\vecl_1|}^{X\phi}  \; (\vecl-\vecl_1) \cdot \vecl \;
\cos 2 (\varphi_{\vecl} -\varphi_{\vecl_1}) + C_{l_1}^{X\phi} \; \vecl_1 \cdot \vecl \;
\cos 2 (\varphi_{\vecl} -\varphi_{|\vecl-\vecl_1|}) \Big]}{
\left[C_{l_1}^{B,\tot} C_{|\vecl-\vecl_1|}^{X,\tot} +C_{|\vecl-\vecl_1|}^{B,\tot} C_{l_1}^{X,\tot} \right]} \, .
\end{eqnarray}
We can define similar filters for the extraction of $\tilde E$ from $(BX)$ or $(EX)$ estimators as well.
In analogous to above, these filters involve the noise contributions in the $(BX)$ and $(EX)$ maps and the
behavior of the function that integrates over the filter in each of the integrals in equations~(\ref{eq:exe}) to
(\ref{eq:bxb}). Since each estimate of $C_l^{\tilde E \tilde E}$ and $C_l^{\tilde B \tilde B}$ comes from two independent
statistics involving $(EX)$ and $(BX)$,  we add the individual signal-to-noise estimates of
the two. This is equivalent to weighing individual noise contributions inversely.

\begin{figure}[t]
\centerline{\psfig{file=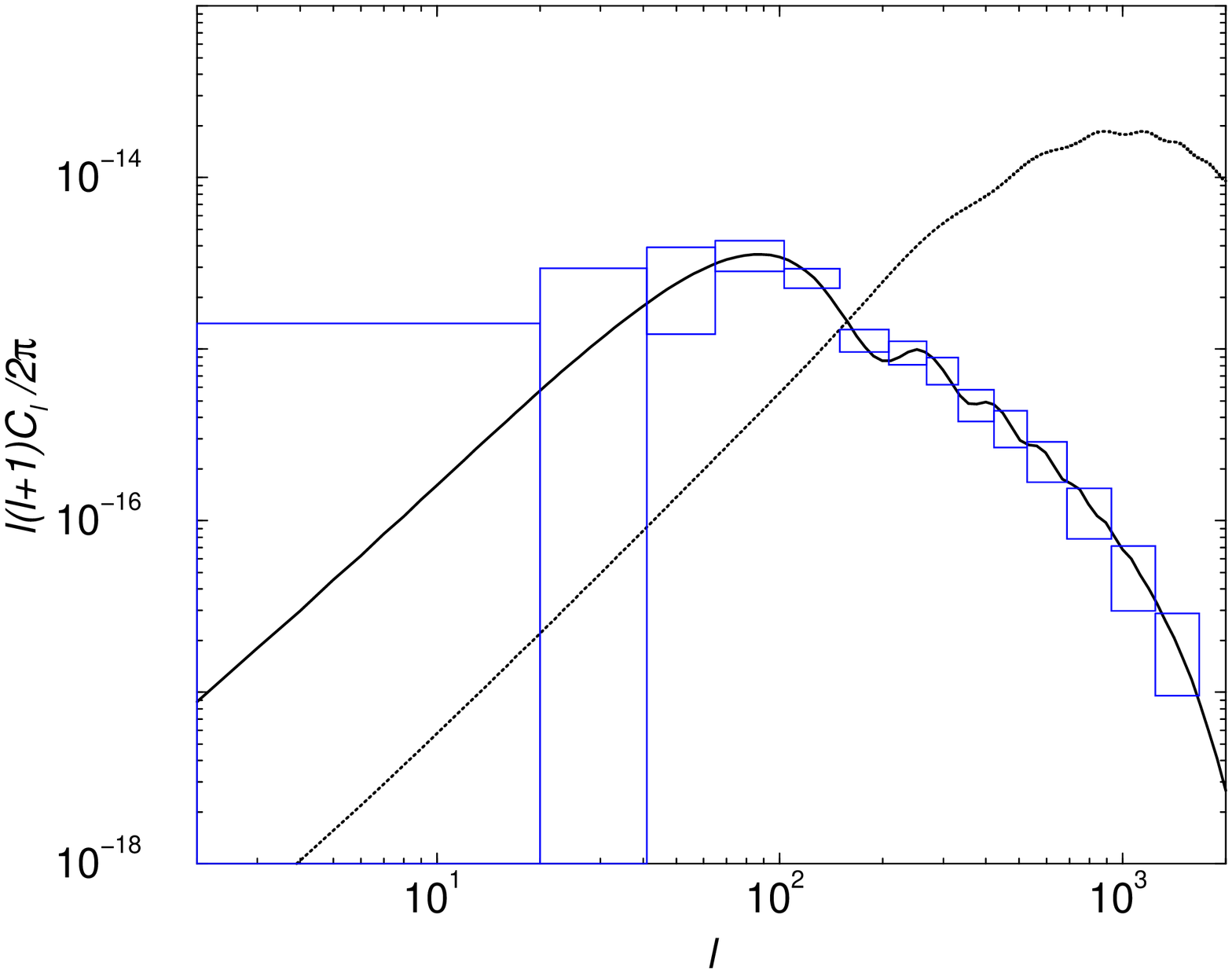,width=3.5in,angle=-90}
\psfig{file=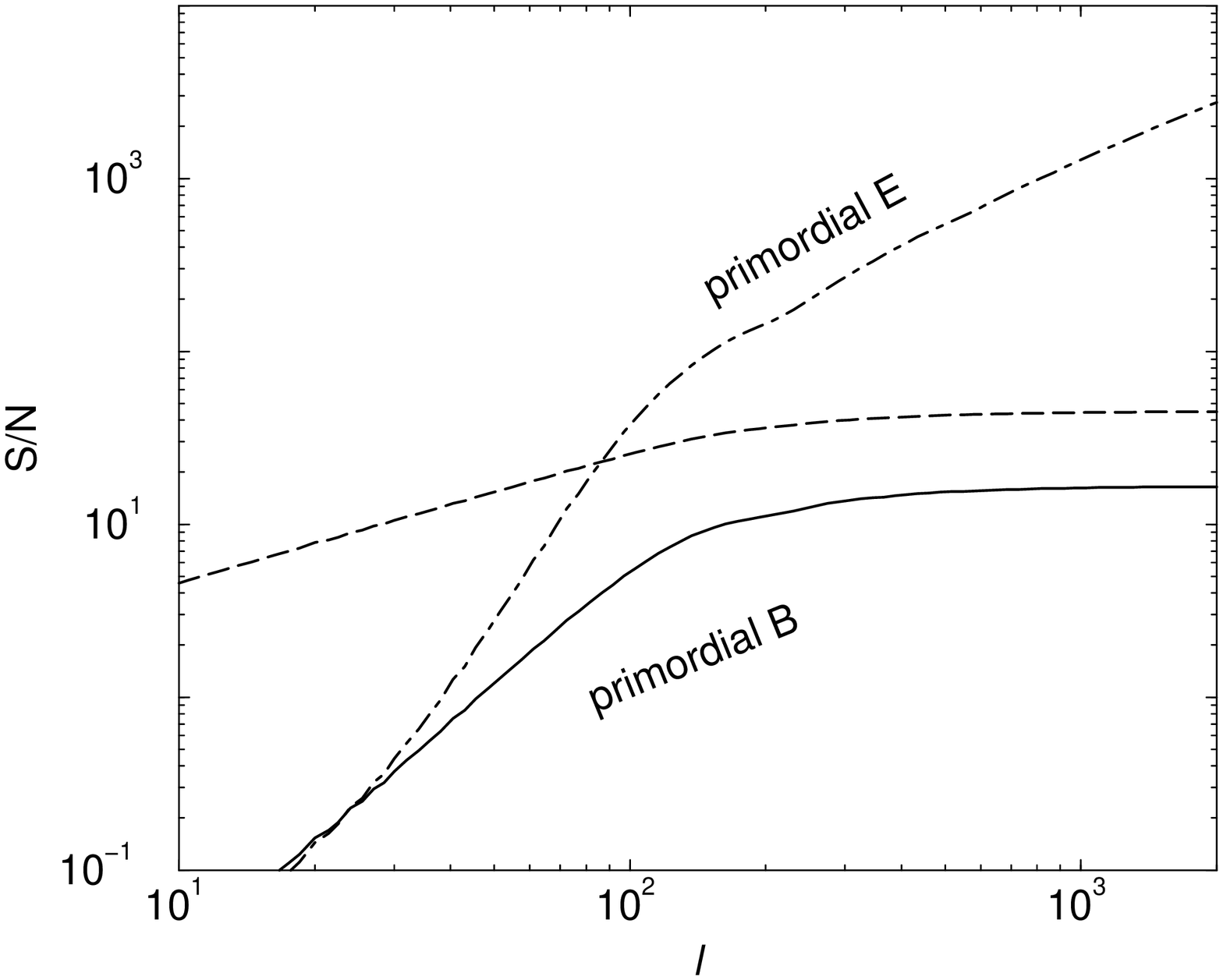,width=3.5in,angle=-90}}
\caption{{\it Left:} The CMB polarization B-modes power spectra. We show the error bars associated with
reconstruction of the primordial power spectrum using no-noise E and B maps and a tracer of the large scale structure
involving lensing potentials. For comparison, the dotted-line shows the contaminant contribution arising from the
transfer of power from primordial E to B-modes via lensing deflections. {\it Right:} The cumulative signal-to-noise
ratios for the reconstruction of the primordial power spectra. The long-dashed line is the signal-to-noise ratio
associated with the detection of B-mode power spectrum, under the assumption that contaminant lensing
contribution can be treated as a source of noise.}
\label{fig:cl}
\end{figure}

\section{Discussion}

We summarize our main results in figure~1 for the extraction of primordial polarization power spectra based on information
related to lensing modification to CMB anisotropies. For the reconstruction, we assume a gravitational-wave contribution to  B-modes
with a tensor-to-scalar ratio of 0.25 and assume no instrumental-noise contributions to  polarization
observations today.  When calculating noise, we include additional secondary noise contribution from lensing to $E$ and $B$ maps.
For simplicity, we also take a tracer field, $X$, that corresponds to lensing potentials themselves. 
Thus, signal-to-noise ratios shown in figure~1 should be considered as optimistic values. Any noise contribution to polarization
observations and a tracer field which is less correlated with $\phi$ will lead to a degradation in the signal-to-noise ratios 
of primordial polarization reconstruction from the ones shown here.

With $X=\phi$, the assumption
here is that one can use a properly cleaned CMB temperature map to reconstruct lensing deflections following the approach in
\cite{Hu01,CooKes02} and apply such a construction to polarization observations to reduce contamination from lensing to
gravitational-wave detection. We include a noise-contribution to the $X$ field, when calculating $C_l^{X,\tot}$, following the
noise calculation for $\phi$ extracted from a no instrumental-noise CMB temperature map in Ref. \cite{Hu01,CooKes02}.
Additionally possibilities for $X$ include a frequency cleaned SZ map, a map of the
convergence from large scale structure lensing observations, the galaxy distribution from surveys such as Sloan but
imaging out to a higher redshift; Since contributions to lensing in CMB comes at redshifts of a few and more,
in general, a tracer field out to redshifts of $\sim$ 3 to 5 will be required to appropriately apply this method. 

As suggested, the primordial polarization construction is more important for B-modes, given that they contain the distinct
signature of gravitational waves, than E-modes. For the extraction of the B-mode power spectrum, we estimate a cumulative
signal-to-noise ratio of $\sim$ 16. The cumulative signal-to-noise for the detection of this contribution,
under the assumption that the contaminant lensing contribution can be treated as a known source of noise, is
of order $\sim$ 40. The suggested reconstruction here leads to a roughly a factor of $\sim$ 2.5 decrease in the associated
signal-to-noise ratio. The reconstruction, however, has the advantage that one does not need to make any assumptions
on the lensing contribution to the power spectrum of B-modes. Relaxing our assumption of a tensor-to-scalar ratio of 0.25,
we find that the suggested extraction can be used to separate primordial B-modes with at least a signal-to-noise ratio of 1
when the tensor-to-scalar ratio is of order 0.01. Though we have not separated individual signal-to-noise
values, the reconstruction of B-modes has less noise with the combination of $(BX)$ statistic correlated with 
a map of B-modes instead of $(EX)$ correlated with B-modes. This is because
the dominant noise associated with E-modes in $(EX)$ limit the reconstruction of $\tilde B$-modes
significantly. The signal-to-noise values for the reconstruction of
$\tilde E$-modes have similar values whether $(EX)$ or $(BX)$ is used. This reconstruction has a cumulative signal-to-noise in excess of $10^3$ both due to its high contribution and the fact that $\tilde E$-mode power spectrum
peak at arcminute scales compared to $\tilde B$-mode power spectrum which peaks at degree scales.

It is clear that the reconstruction analysis can be best done for the extraction of primordial $\tilde E$-modes.
The low signal-to-noise values for $\tilde B$-modes, however, should not discourage from attempting to do analysis like
the one suggested here in upcoming CMB data. The ultimate, and challenging, detection of gravitational wave
signature, which, in principle, gives a handle on the inflationary energy scale requires various statistical
techniques so that this contribution can be identified confusion free. Since lensing contribution 
dominates the B-mode signal, we have exploited its use as a possible way to extract primordial polarization field.
Given the importance of detecting gravitational waves for future cosmological studies, statistics such as the one
discussed are clearly warranted for further studies.

\acknowledgments
This work was supported in part by DoE DE-FG03-92-ER40701 and
a senior research fellowship from the Sherman Fairchild Foundation.

\end{document}